\documentstyle[12pt]{article}

\catcode`\@=11
\long\def\@makefntext#1{ %\parindent 1em
\protect\noindent \hbox to 3.2pt {\hskip-.9pt
$^{{\ninerm\@thefnmark}}$\hfil}#1\hfill} %can be used

\def\thefootnote{\fnsymbol{footnote}}
 \def\@makefnmark{\hbox to 0pt{$^{\@thefnmark}$\hss}}  %original

\def\ps@myheadings{\let\@mkboth\@gobbletwo
\def\@oddhead{\hbox{} %\sl
\rightmark\hfil\ninerm\thepage}
\def\@oddfoot{}\def\@evenhead{\ninerm\thepage\hfil %\sl
\leftmark\hbox{}}\def\@evenfoot{}
\def\sectionmark##1{}\def\subsectionmark##1{}}

\begin{document}

%----------------------------PROCSLA.STY------------------------
\newcommand{\symbolfootnote}{\renewcommand{\thefootnote}
{\fnsymbol{footnote}}}
\renewcommand{\thefootnote}{\fnsymbol{footnote}}
\newcommand{\alphfootnote}
{\setcounter{footnote}{0}
\renewcommand{\thefootnote}{\sevenrm\alph{footnote}}}

%---------------------------------------------------------------
%NEW DEFINED SECTION COMMANDS
\newcounter{sectionc}\newcounter{subsectionc}%
\newcounter{subsubsectionc}
\renewcommand{\section}[1] {\vspace{0.6cm}%
\addtocounter{sectionc}{1}
\setcounter{subsectionc}{0}%
\setcounter{subsubsectionc}{0}\noindent
{\bf\thesectionc. #1}\par\vspace{0.4cm}}
\renewcommand{\subsection}[1] {\vspace{0.6cm}%
\addtocounter{subsectionc}{1}
\setcounter{subsubsectionc}{0}\noindent
{\it\thesectionc.\thesubsectionc. #1}\par\vspace{0.4cm}}
\renewcommand{\subsubsection}[1]{\vspace{0.6cm}%
\addtocounter{subsubsectionc}{1}
\noindent {\rm\thesectionc.\thesubsectionc.\thesubsubsectionc.
#1}\par\vspace{0.4cm}}
\newcommand{\nonumsection}[1] {\vspace{0.6cm}\noindent{\bf #1}
\par\vspace{0.4cm}}

%NEW MACRO TO HANDLE APPENDICES
\newcounter{appendixc}
\newcounter{subappendixc}[appendixc]
\newcounter{subsubappendixc}[subappendixc]
\renewcommand{\thesubappendixc}%
{\Alph{appendixc}.\arabic{subappendixc}}
\renewcommand{\thesubsubappendixc}
{\Alph{appendixc}.\arabic{subappendixc}.%
\arabic{subsubappendixc}}

\renewcommand{\appendix}[1] {\vspace{0.6cm}
        \refstepcounter{appendixc}
        \setcounter{figure}{0}
        \setcounter{table}{0}
        \setcounter{equation}{0}
\renewcommand{\thefigure}{\Alph{appendixc}.\arabic{figure}}
\renewcommand{\thetable}{\Alph{appendixc}.\arabic{table}}
\renewcommand{\theappendixc}{\Alph{appendixc}}
\renewcommand{\theequation}{\Alph{appendixc}.\arabic{equation}}
%\noindent{\bf Appendix \theappendixc. #1}\par\vspace{0.4cm}}
\noindent{\bf Appendix \theappendixc #1}\par\vspace{0.4cm}}
\newcommand{\subappendix}[1] {\vspace{0.6cm}
        \refstepcounter{subappendixc}
\noindent{\bf Appendix \thesubappendixc. #1}\par\vspace{0.4cm}}
\newcommand{\subsubappendix}[1] {\vspace{0.6cm}
        \refstepcounter{subsubappendixc}
        \noindent{\it Appendix \thesubsubappendixc. #1}
\par\vspace{0.4cm}}

%----------------------------------------------------------------
%MARCO FOR ABSTRACT BLOCK
\def\abstracts#1{{
\centering{\begin{minipage}{30pc}\tenrm%
\baselineskip=12pt\noindent
\centerline{\tenrm ABSTRACT}\vspace{0.3cm}
\parindent=0pt #1
\end{minipage} }\par}}

%---------------------------------------------------------------
%NEW MACRO FOR BIBLIOGRAPHY
\newcommand{\bibit}{\it}
\newcommand{\bibbf}{\bf}
\renewenvironment{thebibliography}[1]
{\begin{list}{\arabic{enumi}.}
{\usecounter{enumi}\setlength{\parsep}{0pt}
%1.25cm IS STRICTLY FOR PROCSLA.TEX ONLY
\setlength{\leftmargin 1.25cm}{\rightmargin 0pt}
%0.52cm IS FOR NEW DATA FILES
%\setlength{\leftmargin 0.52cm}{\rightmargin 0pt}
 \setlength{\itemsep}{0pt} \settowidth
{\labelwidth}{#1.}\sloppy}}{\end{list}}

%----------------------------------------------------------
%FOLLOWING THREE COMMANDS ARE FOR 'LIST' COMMAND.
\topsep=0in\parsep=0in\itemsep=0in
\parindent=1.5pc

%LIST ENVIRONMENTS
\newcounter{itemlistc}
\newcounter{romanlistc}
\newcounter{alphlistc}
\newcounter{arabiclistc}
\newenvironment{itemlist}
    {\setcounter{itemlistc}{0}
 \begin{list}{$\bullet$}
{\usecounter{itemlistc}
 \setlength{\parsep}{0pt}
 \setlength{\itemsep}{0pt}}}{\end{list}}

\newenvironment{romanlist}
{\setcounter{romanlistc}{0}
 \begin{list}{$($\roman{romanlistc}$)$}
{\usecounter{romanlistc}
 \setlength{\parsep}{0pt}
 \setlength{\itemsep}{0pt}}}{\end{list}}

\newenvironment{alphlist}
{\setcounter{alphlistc}{0}
 \begin{list}{$($\alph{alphlistc}$)$}
{\usecounter{alphlistc}
 \setlength{\parsep}{0pt}
 \setlength{\itemsep}{0pt}}}{\end{list}}

\newenvironment{arabiclist}
{\setcounter{arabiclistc}{0}
 \begin{list}{\arabic{arabiclistc}}
{\usecounter{arabiclistc}
 \setlength{\parsep}{0pt}
 \setlength{\itemsep}{0pt}}}{\end{list}}

%--------------------------------------------------------------
%FIGURE CAPTION
\newcommand{\fcaption}[1]{
        \refstepcounter{figure}
        \setbox\@tempboxa = \hbox{\tenrm Fig.~\thefigure. #1}
        \ifdim \wd\@tempboxa > 6in
           {\begin{center}
\parbox{6in}{\tenrm\baselineskip=12pt Fig.~\thefigure. #1 }
            \end{center}}
        \else
             {\begin{center}
             {\tenrm Fig.~\thefigure. #1}
              \end{center}}
        \fi}

%TABLE CAPTION
\newcommand{\tcaption}[1]{
        \refstepcounter{table}
        \setbox\@tempboxa = \hbox{\tenrm Table~\thetable. #1}
        \ifdim \wd\@tempboxa > 6in
           {\begin{center}
\parbox{6in}{\tenrm\baselineskip=12pt Table~\thetable. #1 }
            \end{center}}
        \else
             {\begin{center}
             {\tenrm Table~\thetable. #1}
              \end{center}}
        \fi}

%---------------------------------------------------------------
%ACKNOWLEDGEMENT: this portion is from John Hershberger
\def\@citex[#1]#2{\if@filesw\immediate\write\@auxout
{\string\citation{#2}}\fi
\def\@citea{}\@cite{\@for\@citeb:=#2\do
{\@citea\def\@citea{,}\@ifundefined
{b@\@citeb}{{\bf ?}\@warning
{Citation `\@citeb' on page \thepage \space undefined}}
{\csname b@\@citeb\endcsname}}}{#1}}

\newif\if@cghi
\def\cite{\@cghitrue\@ifnextchar [{\@tempswatrue
\@citex}{\@tempswafalse\@citex[]}}
\def\citelow{\@cghifalse\@ifnextchar [{\@tempswatrue
\@citex}{\@tempswafalse\@citex[]}}
\def\@cite#1#2{{$\null^{#1}$\if@tempswa\typeout
{IJCGA warning: optional citation argument
ignored: `#2'} \fi}}
\newcommand{\citeup}{\cite}

%------------------------------------------------------------
%FOR FNSYMBOL FOOTNOTE AND ALPH{FOOTNOTE}
\def\fnm#1{$^{\mbox{\scriptsize #1}}$}
\def\fnt#1#2{\footnotetext{\kern-.3em
{$^{\mbox{\sevenrm #1}}$}{#2}}}

%------------------------------------------------------------
\font\twelvebf=cmbx10 scaled\magstep 1
\font\twelverm=cmr10 scaled\magstep 1
\font\twelveit=cmti10 scaled\magstep 1
\font\elevenbfit=cmbxti10 scaled\magstephalf
\font\elevenbf=cmbx10 scaled\magstephalf
\font\elevenrm=cmr10 scaled\magstephalf
\font\elevenit=cmti10 scaled\magstephalf
\font\bfit=cmbxti10
\font\tenbf=cmbx10
\font\tenrm=cmr10
\font\tenit=cmti10
\font\ninebf=cmbx9
\font\ninerm=cmr9
\font\nineit=cmti9
\font\eightbf=cmbx8
\font\eightrm=cmr8
\font\eightit=cmti8

%%%%%%%%%%%%%%%%%%%%%%%%%%%%%%%%%%%%%%%%%%%%%%%%%%%%%%%%%%%%%%%%%%%%%%%%
%----------------------START OF DATA FILE--------------------
\newcommand{\be}{\begin{equation}}
\newcommand{\ee}{\end{equation}}
\newcommand{\bea}{\begin{eqnarray}}
\newcommand{\eea}{\end{eqnarray}}
\newcommand{\bd}{\begin{displaymath}}
\newcommand{\ed}{\end{displaymath}}
\newcommand{\rar}{\rightarrow}
\newcommand{\lar}{\leftarrow}
\newcommand{\half}{\frac{1}{2}}
\newcommand{\lnb}{\mbox{\Large$\left(\right.$}}
\newcommand{\rnb}{\mbox{\Large$\left.\right)$}}
\newcommand{\lsb}{\mbox{\LARGE$\left[\right.$}}
\newcommand{\rsb}{\mbox{\LARGE$\left.\right]$}}
\newcommand{\lcb}{\mbox{\Large$\left\{\right.$}}
\newcommand{\rcb}{\mbox{\Large$\left.\right\}$}}
\newcommand{\LCB}{\mbox{\LARGE$\left\{\right.$}}
\newcommand{\RCB}{\mbox{\LARGE$\left.\right\}$}}
\newcommand{\lla}{\mbox{\LARGE$\left\langle\right.$}}
\newcommand{\rra}{\mbox{\LARGE$\left.\right\rangle$}}

\centerline{\tenbf FERMION NUMBER VIOLATION AND}
\baselineskip=22pt
\centerline{\tenbf A TWO-DIMENSIONAL HIGGS MODEL}
\baselineskip=16pt
\vspace{0.8cm}
\centerline{\tenrm I. MONTVAY}
\baselineskip=13pt
\centerline{\tenit Deutsches Elektronen-Synchrotron DESY, Notkestr. 85}
\baselineskip=12pt
\centerline{\tenit D-22603 Hamburg, FRG}
\vspace{0.9cm}
\abstracts{ The investigation of topological properties of the gauge
 field in a two-dimensional Higgs model can help in understanding
 anomalous fermion number violation. }

%%%%%%%%%%%%%%%%%%%%%%%%%%%%%%%%%%%%%%%%%%%%%%%%%%%%%%%%%%%%%%%%%%%%%%

\vspace{0.5cm}
\twelverm
\baselineskip=14pt
\section{ A simple model for fermion number violation }
 Fermion number, which is the sum of baryon number and lepton number
 ($B+L$), is not conserved in the Standard Model \cite{THOOFT}.
 This is due to the anomaly in the fermion current.
 The lattice formulation of the anomalous fermion number
 non-conservation is problematic \cite{BANKS}, because it has to do with
 the chiral $\rm SU(2)_L \otimes U(1)_Y$ gauge couplings and, as is well
 known, there is a difficulty with chiral gauge theories on the lattice
 (see, for instance, the reviews \cite{TSUKPR,PETCHER}).
 There is, however, an approximation of the electroweak sector of the
 standard model which can be studied with standard lattice techniques,
 namely the limit when the
 $\rm SU(3)_{colour} \otimes U(1)_{hypercharge}$ gauge couplings are
 zero \cite{AMSTPR,JOHOPR}.

 A simple prototype model is the standard $\rm SU(2)_L$ Higgs model
 coupled to an even number $2N_f$ of fermion doublets.
 One can take, for simplicity, $N_f=1$ but the extension to $N_f > 1$
 is straightforward.
 The lattice action can be written in terms of the fermion doublet
 fields $\psi_{(1,2)x}$ with an off-diagonal Majorana mass and
 Majorana-like Wilson term.
 It is, however, technically more convenient to consider a Dirac-like
 form with $\psi \equiv \psi_1$ and the mirror fermion field $\chi$
 defined by
\be \label{eq01}
\chi_x \equiv \epsilon^{-1} C \overline{\psi}_{2x}^T \ ,
\hspace{2em}
\overline{\chi}_x \equiv \psi_{2x}^T \epsilon C  \ .
\ee
 with the charge conjugation matrix $C$ and $\epsilon$ acting in isospin
 space.
 In terms of $\psi$ and $\chi$ one obtains the mirror fermion action for
 chiral gauge fields, which is  well suited for studying the physically
 relevant phase with broken symmetry.

 Before doing numerical simulations in the four dimensional
 $\rm SU(2)_L$ gauge model for anomalous fermion number violation, it
 is useful to study a simple U(1) toy model in two dimensions, which has
 often been studied in this context (see e.~g. \cite{BOCSHA}).

 The lattice action depending on the compact U(1) gauge field
 $U_{x\mu}=\exp(iA_\mu(x))$, $(\mu=1,2)$ and, for simplicity, fixed
 length Higgs scalar field $\phi(x),\; |\phi(x)|=1$ can be
 written as
\be \label{eq02}
S = \beta \sum_x \sum_{\mu=1, \nu=2}
[1 - cos(F_{\mu\nu}(x))]
-2\kappa \sum_x \sum_{\mu=1}^2 \phi^*(x+\hat{\mu})U_{x\mu}\phi(x) \ ,
\ee
 where the lattice gauge field strength is defined for $\mu,\nu=1,2$ as
\be \label{eq03}
F_{\mu\nu}(x) =
A_\nu(x+\hat{\mu})-A_\nu(x)-A_\mu(x+\hat{\nu})+A_\mu(x) \ .
\ee
 Real angular variables $-\pi < \theta_{x\mu} \leq \pi$ on the links
 can be introduced by
\be \label{eq04}
U_{x\mu} \equiv \exp(i\theta_{x\mu}) \ , \hspace{2em}
\theta_{x\mu} = A_\mu(x)-2\pi \cdot \mbox{\rm NINT}(A_\mu(x)/2\pi) \ ,
\ee
 where NINT() denotes nearest integer.

 Fermions in this two dimensional model are introduced in the
 mirror fermion basis $(\psi,\chi)$.
 The U(1) gauge field is only coupled to $\psi_L$ and $\chi_R$.
 Hence, using the notations $P_{L,R}=(1 \mp \gamma_5)/2$ and
\be \label{eq05}
U_{(L,R)x\mu}=P_{(L,R)} U_{x\mu}+P_{(R,L)} \ ,
\ee
 the fermionic part of the action is:
\bd
S_{fermion} = \sum_x \LCB
  \mu_0 \left[ (\overline{\chi}_x\psi_x)
+ (\overline{\psi}_x\chi_x) \right]
- \half \sum_{\mu = \pm 1}^{\pm 2} \lsb
  (\overline{\psi}_{x+\hat{\mu}} \gamma_\mu U_{Lx\mu} \psi_x)
+ (\overline{\chi}_{x+\hat{\mu}} \gamma_\mu U_{Rx\mu} \chi_x)
\ed
\be \label{eq06}
 - r \lnb
  (\overline{\chi}_x\psi_x)
- (\overline{\chi}_{x+\hat{\mu}} U_{Lx\mu}\psi_x)
+ (\overline{\psi}_x\chi_x)
- (\overline{\psi}_{x+\hat{\mu}} U_{Rx\mu} \chi_x) \rnb \rsb \RCB \ .
\ee
 Here $r$ is the Wilson parameter of the fermions and $\mu_0$ the bare
 fermion mass.
 Possible Yukawa couplings $G_\psi$ and $G_\chi$ of the $\psi$-,
 respectively, $\chi$-fields are set here equal to zero for simplicity.

 The fermion number current in the four dimensional $\rm SU(2)_L$ gauge
 model is the difference of the number of fermions ($\psi$) and
 number of mirror fermions ($\chi$).
 The corresponding current in the two dimensional U(1) model,
 for $\mu=1,2$, is:
\bd
J_{x\mu} \equiv \frac{1}{2} \left[
(\overline{\psi}_{x+\hat{\mu}} \gamma_\mu U_{Lx\mu} \psi_x)
+(\overline{\psi}_x \gamma_\mu U_{Lx\mu}^* \psi_{x+\hat{\mu}})
\right.
\ed
\be \label{eq07}
\left.
-(\overline{\chi}_{x+\hat{\mu}} \gamma_\mu U_{Rx\mu} \chi_x)
-(\overline{\chi}_x \gamma_\mu U_{Rx\mu}^* \chi_{x+\hat{\mu}})
\right] \ ,
\ee
 where the two dimensional $\gamma$-matrices are defined as
 $\gamma_1=\sigma_1,\; \gamma_2=\sigma_2,\; \gamma_5=\sigma_3$.
 Another alternative choice for the fermion fields in this model is
 to use the reshuffled combinations
\be \label{eq08}
\psi_{Cx} \equiv \psi_{Lx}+\chi_{Rx} \ , \hspace{1em}
\psi_{Nx} \equiv \chi_{Lx}+\psi_{Rx} \ .
\ee
 Then the gauge field is coupled only to the {\em ``charged''} field
 $\psi_C$, the other field $\psi_N$ is {\em ``neutral''}.
 In terms of these the current $J_{x\mu}$ is a combination of the
 axialvector currents:
\bd
J_{x\mu} \equiv \frac{1}{2} \left[
-(\overline{\psi}_{C,x+\hat{\mu}} \gamma_\mu \gamma_5 U_{x\mu}
  \psi_{Cx})
-(\overline{\psi}_{Cx}
  \gamma_\mu \gamma_5 U_{x\mu}^* \psi_{C,x+\hat{\mu}})
\right.
\ed
\be \label{eq09}
\left.
+(\overline{\psi}_{N,x+\hat{\mu}} \gamma_\mu \gamma_5 \psi_{Nx})
+(\overline{\psi}_{Nx} \gamma_\mu \gamma_5 \psi_{N,x+\hat{\mu}})
\right] \ .
\ee

 The anomalous Ward-Takahashi identity for the fermion current
 $J_{x\mu}$ can be derived in the usual way on a small and smooth
 background gauge field.
 In the continuum limit $a \rar 0$ one obtains, with the backward
 lattice derivative $\Delta^b_\mu$,
\be \label{eq10}
\langle \Delta^b_\mu J_{x\mu} \rangle
= 4 {\cal K}(r,\mu_0) \epsilon_{\mu\nu} \partial_\mu A_\nu(x) + O(a^3)
\ .
\ee
 The lattice integral $\cal K$ is given by
\be \label{eq11}
{\cal K}(r,\mu_0) \equiv \frac{1}{(2\pi)^2} \int_{-\pi}^\pi
\frac{\mu_k \cos k_1 \cos k_2}{(\bar{k}^2 + \mu_k^2)^2} \cdot
[ r \sum_{\alpha=1}^2 \bar{k}_\alpha^2/\cos k_\alpha - \mu_k ] d^2 k\ ,
\ee
 and the notations are
\be \label{eq12}
\mu_k = \mu_0 + \frac{r}{2} \hat{k}^2 \ ,  \;
\bar{k}_\mu = \sin k_\mu \ ,               \;
\hat{k}_\mu = 2\sin \frac{k_\mu}{2} \ .
\ee
 For vanishing bare fermion mass $\mu_0=0$ one can prove (see e.~g.
 \cite{KARSMI,SEISTA}) that the integral ${\cal K}$ is
\be \label{eq13}
{\cal K}(r,0) = \frac{1}{4\pi} \hspace{2em}
({\rm independently\; from}\; r) \ .
\ee
 Therefore we get the anomaly equation
\be \label{eq14}
\langle \partial_\mu J_\mu(x) \rangle =
\lim_{a \to 0} \langle \Delta^b_\mu J_{x\mu} \rangle a^{-2}
= \frac{1}{2\pi} \epsilon_{\mu\nu} F_{\mu\nu}(x) = 2 q(x) \ .
\ee
 Here $q(x)$ is the density of the topological charge
\be \label{eq15}
q(x)= \frac{1}{4\pi} \epsilon_{\mu\nu} F_{\mu\nu}(x) \ .
\ee

 Since according to (\ref{eq14}) the non-conservation of the fermion
 current $J_{x\mu}$ is proportional to the topological charge
 density, the first step in understanding the anomaly on the lattice
 is to understand the topological features of the two dimensional U(1)
 lattice gauge fields.
 Some recent results of an exploratory numerical study in this model
 are given in the Proceedings of Lattice '93 Conference \cite{DALLAS}.

%%%%%%%%%%%%%%%%%%%%%%%%%%%%%%%%%%%%%%%%%%%%%%%%%%%%%%%%%%%%%%%%%%%%%%

\end{document}